\documentclass{aa}  

\usepackage{graphicx}
\usepackage{txfonts}
\usepackage[colorlinks=true,citecolor=blue, linkcolor=blue,
     filecolor=blue,
     urlcolor = blue]{hyperref}
\usepackage{multirow}
\usepackage{tablefootnote}
\usepackage{float}

\begin{document}

   \title{Flares detected in ALMA single-dish images of the Sun}

   \subtitle{}

   \author{I. Skoki\'c
          \inst{1}
          \and
		  A.~O. Benz\inst{2}\fnmsep\inst{3}
          \and
          R. Braj\v{s}a\inst{1}          
          \and
          D. Sudar\inst{1}
          \and
          F. Matkovi\'{c}\inst{1}
          \and
          M. B\'{a}rta\inst{4}
          }

   \institute{Hvar Observatory, Faculty of Geodesy, University of Zagreb,
              Ka\v{c}i\'ceva 26, HR-10000 Zagreb, Croatia\\
              \email{ivica.skokic@gmail.com}
         \and
             University of Applied Sciences and Arts Northwestern Switzerland, Bahnhofstrasse 6, CH-5210 Windisch, Switzerland 
         \and
              Institute for Particle Physics and Astrophysics, ETH Z\"urich, CH-8093 Z\"urich, Switzerland
\and
              Astronomical Institute of the Czech Academy of Sciences, Fri\v{c}ova 298, CZ-25165 Ond\v{r}ejov, Czech Republic              
             }

   \date{Received -; accepted -}

 
  \abstract
   {The millimeter and submillimeter radiation of solar flares is poorly understood. Without spatial resolution, it cannot be compared easily to flare emissions in other wavelengths. The Atacama Large Millimeter-submillimeter Array (ALMA) offers sufficient resolution for the first time. However, used as an interferometer, its field of view is smaller than an active region and ALMA cannot observe on demand when a flare occurs. }
   {We use readily available large scale single-dish ALMA observations of solar millimeter flares and compare them to well-known features observed in other wavelengths. The properties of these other flare emissions, correlating in space and time, may then be used to interpret the millimeter brightenings and vice versa. The aim is to obtain reliable associations, limited by the time and space resolution of single-dish observations.}
  {Ordinary  interferometric ALMA observations require single-dish images of the full Sun for calibration. We collected such observations at 3 mm and 1 mm and searched for millimeter brightenings during times given in a flare catalog. }
   {All of the flares left a signature in millimeter waves. We found five events with 9 or more images that can be used for comparison in time and space. The millimeter brightenings are associated with a variety of flare features in cool (H$\alpha$, 304 \AA), intermediate (171 \AA), and hot (94 \AA) lines. In several cases, the millimeter brightening peaked at the footpoint of a hot flare loop. In other cases the peak of the millimeter brightening coincided with the top or footpoint of an active H$\alpha$ filament. We found correlations also with post-flare loops and tops of a hot loop. In some images the millimeter radiation peaked at locations, where no feature in the selected lines was found.}
   {The wide field of view provided by the single-dish observations allowed for completely overviewing the flare activity in millimeter waves for the first time. The associated phenomena often changed during the flare in type and location. The variety of phenomena detected in these millimeter observations may explain the sometimes bewildering behavior of millimeter flare emissions observed previously without spatial resolution. }

   \keywords{Sun: chromosphere --
                Sun: flares -- Sun: filaments -- Sun: radio radiation 
               }

   \maketitle
%

\section{Introduction}

Brightenings during solar flares have been reported over the full electromagnetic spectrum from gamma-rays to decameter radio waves  \citep{Benz2017LRSP}. They are caused by various drivers and mechanisms, but their interpretation remains fragmentary. Little is known about the range from 100 GHz to 10 THz (mid infrared).  Contrary to the flare emission below 100 GHz \citep{Bastian1998}, the spectrum above sometimes increases with frequency, rising the possibility of a "THz-component" \citep{Kaufmann2004}. The millimeter and sub-millimeter flare observations have been reviewed by \citet{Krucker2013}. The >100 GHz emission correlates best, but not completely with HXR and gamma-rays, sometimes with chromospheric lines. The correlation is often different in the pre-flare, impulsive, and even post-impulsive phases of a flare and in different flares. In one case, where this was reliably measured, the position of the 210 GHz emission was on one of the flare ribbons \citep{Luthi2004AA}. \citet{Krucker2013} list ten mechanisms that have been proposed to interpret the THz-component. They remark that there may indeed be more than one explanation.

Interferometric observations using the Berkeley-Illinois-Maryland Array (BIMA) at 86 GHz found that the millimeter emission of a solar flare exhibits impulsive and gradual phases, both often observed in the same flare \citep{kundu1996radio}. High spatial resolution images of solar flares at millimeter wavelengths obtained by BIMA showed that most of the gradual phase millimeter flux came from the top of a flaring loop, with some contribution from the footpoints \citep{Silva1996ApJS,Silva1996,Silva1997,Silva1998}. The millimeter emission from the gradual phase was likely due to thermal bremsstrahlung from the soft X-ray-emitting hot plasma, while gyrosynchrotron radiation was suggested as the main mechanism for the impulsive phase. \citet{Raulin1999} et al. found indications of two different electron populations in the impulsive phase, one responsible for HXR/microwave emission, the other for millimeter emission. \citet{Kundu2000} analyzed two solar flares simultaneously observed at 17, 34, and 86 GHz and reported evidence of bipolar (looplike) structures.

The major limit of the previous >100 GHz observations was spatial resolution, impeding the combination with observations at other wavelengths. The Atacama Large Millimeter-submillimeter Array (ALMA) has a great potential to mitigate this instrumental gap \citep{Wedemeyer2016SSR, Bastian2018}. Since ALMA is not a solar dedicated instrument, however, sporadic flares are difficult to observe. 

For this reason, solar ALMA observations in the past have focused on stationary phenomena such as radius, center-to-limb and center -to-pole brightnening measurements \citep{Alissandrakis2017, Selhorst2019, Sudar2019, Sudar2019CEAB, Menezes2021, Menezes2022},  identification and analysis of various stationary structures in full disk \citep{Brajsa2018AA, Skokic2020CEAB} and interferometric ALMA images \citep{Iwai2017, Bastian2017, Nindos2018, Jafarzadeh2019, Rodger2019, Loukitcheva2019, Molnar2019, Shimojo2020, Wedemeyer2020, Brajsa2020CEAB, Brajsa2021}.

Recent research extends ALMA results to stationary prominences/filaments \citep{Heinzel2022, daSilvaSantos2022, Labrosse2022},  dynamical phenomena such as oscillations in the quiet Sun  \citep{Patsourakos2020, Jafarzadeh2021, Chai2022} and transients \citep{Nindos2020AA,daSilvaSantos2020,Eklund2020}, and even to submillimeter structures \citep{Alissandrakis2022}.

The only known flare-related ALMA observation to date is by \citet{Shimizu2021ApJ}, who observed an active region with ALMA in interferometric mode at 100 GHz. The field of view was 60\arcsec and the resolution  5.\arcsec0 x 3.\arcsec9. They report a microflare, which correlated in time with Si IV line emission ($\sim 80\,000$~K) and in space with the footpoint of an SXR loop in the outskirts of the active region.

Here we report on the first solar flares observed by ALMA with complete spatial coverage. The aim of this paper is to analyze the temporal and spatial evolution of five flares observed by ALMA in single-dish mode. We compare images and  time profiles with data from other wavelengths to identify counterparts and to provide clues for the mechanisms responsible for flare emission at millimeter wavelengths.


\section{Data and methodology}

\begin{table}
\caption{ALMA observation details. In the period column, start times of the first and last ALMA observation of that period are given. Number of observations in the period is given in the column denoted by N.}             
\label{tabALMAobs}      
\centering                          
\begin{tabular}{c c c c c c}        
\hline\hline                 
Date & Period UT & Project code & Freq. & N\\
    &   &   &GHz& \\    
\hline                        
   2017-04-23 & 14:06 -- 16:30 & 2016.1.01129.S & 230 &  9 \\
   2017-04-26 & 14:17 -- 16:19 & 2016.1.00070.S & 107 & 11 \\
   2018-04-03 & 13:47 -- 17:39 & 2017.1.00072.S &  95 & 23 \\
   2018-04-19 & 15:26 -- 17:47 & 2017.1.01138.S &  95 & 14 \\
   2018-12-15 & 13:08 -- 15:12 & 2018.1.01879.S &  95 & 10 \\
\hline                                  

\end{tabular}
\end{table}

Observations of the Sun with the ALMA interferometer require total power full-disk  images for absolute calibration \citep{White2017, Shimojo2017}. These full-disk solar images are publicly available from the ALMA Science Archive\footnote{\url{https://almascience.eso.org}}. We limited the data set to ALMA bands 3 (100~GHz, $\lambda$=3~mm) and 6 (240~GHz, $\lambda$=1.2~mm) because other bands were either not public at the time of retrieval or only a small number of images are available. The final data set consisted of 372 images taken between 2016 Dec 21 and 2019 Apr 13. They amount to a total on-source observing time of about 40 hours. The images were then converted to a helioprojective coordinate system and rotated to the Sun's north pole pointing upward \citep{skokic2019seg}. We improved the alignment in all images by limb fitting and double-checked on compact structures. Next, the images were scaled in intensity so that the brightness temperature of the quiet Sun region nearest to the solar center is equal to 7300~K (5900~K) in band 3 (6), as suggested by \citet{White2017}. \citet{Alissandrakis2020AA} reported that in band 6 the value was most probably underestimated by $\sim$440~K, but we did not include this correction in the present analysis. Finally, the center-to-limb brightness variation was removed by the procedure described in \citet{Sudar2019}.

Depending on the band, it takes ALMA between 10 and 16 minutes to obtain one full-disk image of the Sun. This is the time the antenna needs to scan the Sun and perform additional calibration scans limiting the maximum cadence. The actual scanning of the Sun without calibration takes 301.4~s in band 3 and 581.4~s in band 6, thus defining the temporal uncertainty of the images. We transferred the DATE-OBS time in the FITS headers of the ALMA images referring to the beginning of the scan to the middle of the scan.

\begin{table}
\caption{Parameters of the observed flare events retrieved from the HEK database. Class column denotes the GOES flare class, and HPC column lists helioprojective coordinates of the flare in arcseconds at peak time.}             
\label{tabFlareList}      
\centering                          
\begin{tabular}{c c c c c c c}        
\hline\hline                 
Date & Start & Peak & End &GOES &HPC  \\
    &   &   &   &Class &x, y  \\    
\hline                        
   2017-04-23 & 15:47 & 15:57 & 16:12 & B1.8 & 15, 470 \\
   2017-04-26 & 15:28 & 15:38 & 15:50 & B3.4 & 350, 300 \\
   2018-04-03 & 14:13 & 14:17 & 14:32 & B1.2 & -275, -50\\
   2018-04-19 & 15:57 & 16:29 & 17:25 & - & 750, -50 \\
   2018-12-15 & 13:40 & 13:53 & 14:15 & B1.0 & -640, 215\\
\hline                                   
\end{tabular}
\end{table}
%

   \begin{figure}[] 
   \centering
   \includegraphics[width=9.2 cm]{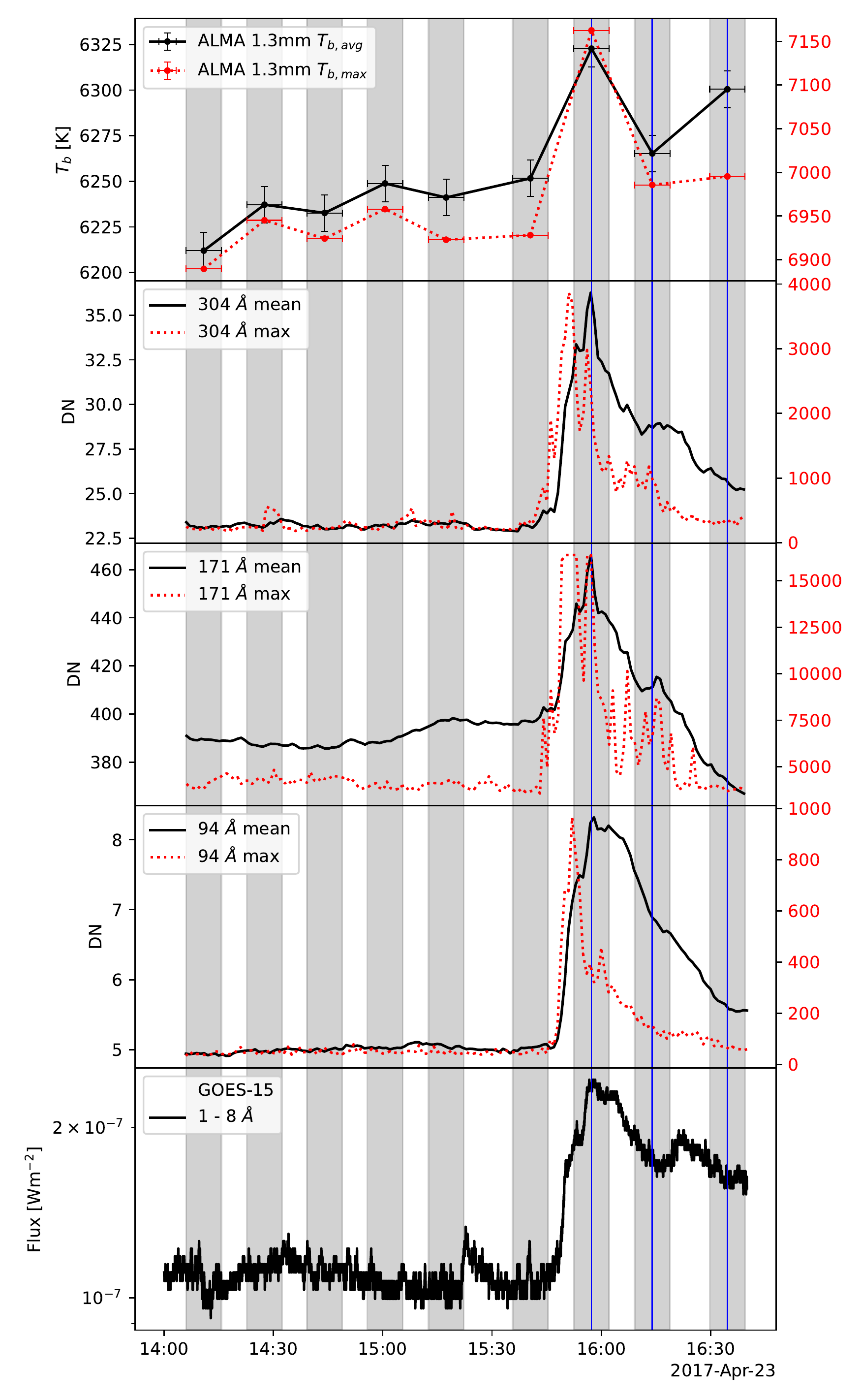}
   \caption{Intensity profiles for the SOL2017-04-23 flare. Gray shaded areas
   represent time intervals during which an ALMA image was obtained. Black curves show intensities spatially averaged over the flare region, and red curves show peak
   intensities within the flare region. The scale of the red curves is given on the right axis, in the same units as the black curves given on the left axis}. Vertical blue lines denote instants
   that are shown in cutout images and further analyzed.
   The same representation is used in all subsequent intensity profiles.
              \label{fig-lc-20170423}%
    \end{figure}
%

   \begin{figure*}[htb!]
   \centering
   \includegraphics[width=18.5cm]{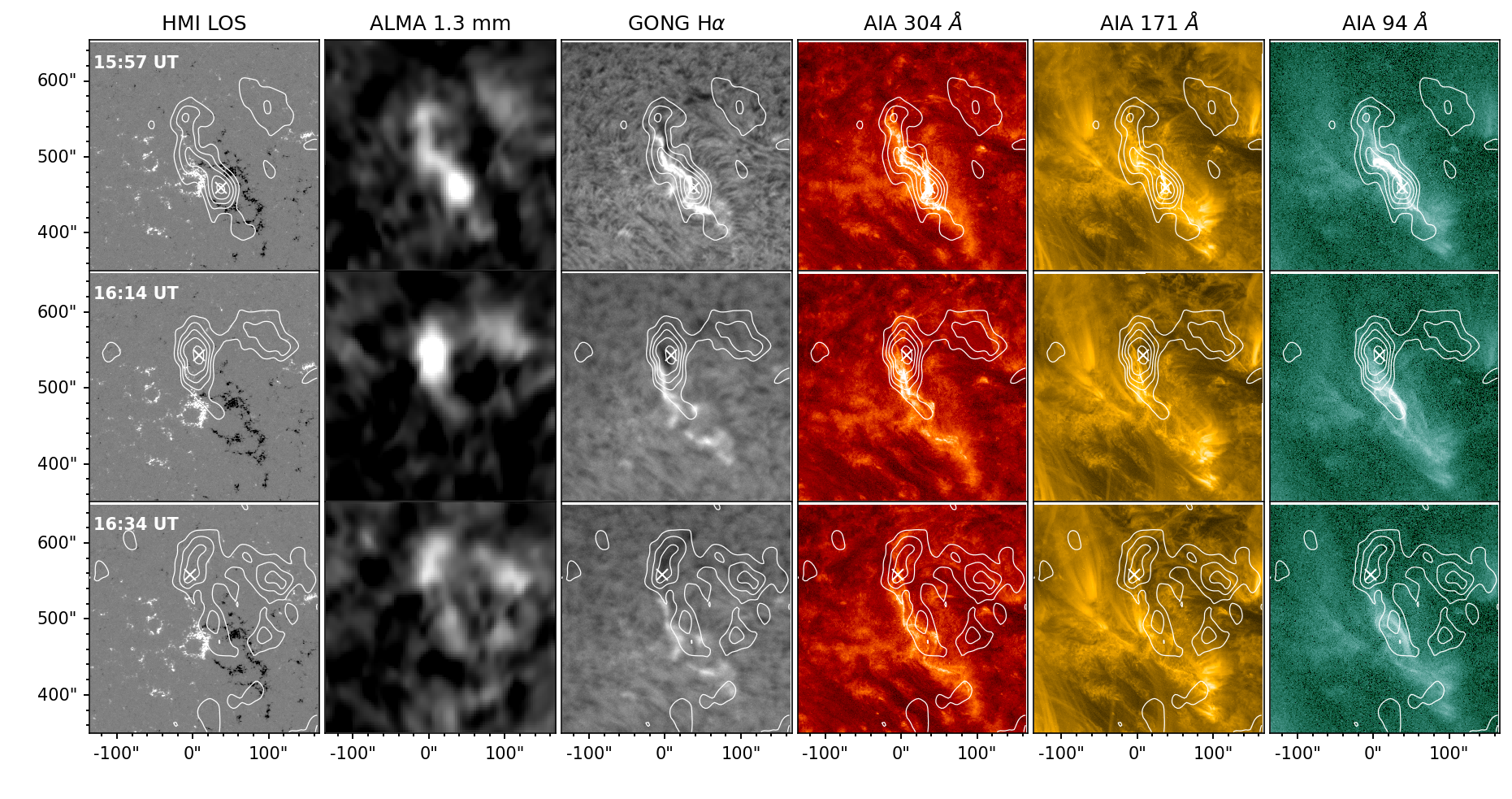}
   \caption{SOL2017-04-23 flare. The ALMA images are difference images between the designated time frame and the base pre-flare frame. The other images are regular full-scale. ALMA contours outline five levels equidistantly within the range of 100 - 500~K. The peak ALMA brightness in the flaring region, is marked with a white "x". The intensity was clipped in each image to better show structures of interest. }
              \label{fig20170423}%
    \end{figure*}

The observation times of the ALMA images were used to search for coinciding flare events in the Heliophysics Events Knowledgebase (HEK\footnote{\url{https://www.lmsal.com/hek/}}). The search returned a list of nearly hundred entries. However, many of them were related to the same flare event reported by different observatories or detection methods. These multiple entries were filtered out, along with short duration events that occurred between two consecutive ALMA observations and events in which ALMA images were of poor quality due to the presence of scan artifacts or other effects. A total of four flare events were singled out. Additionally, one event was manually found that was visually detected as a small brightening in the ALMA data, but was not present in the HEK flare catalog. Details of ALMA observations for all five selected events are listed in Table~\ref{tabALMAobs}. Flare properties obtained from the HEK catalog are listed in Table~\ref{tabFlareList}.

ALMA images were compared with filtergrams in extreme ultraviolet (EUV) from the Atmospheric Imaging Assembly (AIA, \citealt{Lemen2012}) and magnetograms from the Helioseismic and Magnetic Imager (HMI, \citealt{Scherrer2012}) on the Solar Dynamics Observatory (SDO, \citealt{Pesnell2012}). These complementary data were obtained from the Joint Science Operations Center (JSOC\footnote{\url{http://jsoc.stanford.edu}}). We limited the analysis to AIA 94~{\AA}  (flaring regions, characteristic temperature of $T=6\times10^6$ ~K), 171~{\AA} (quiet corona and upper transition region,  $T=6\times10^5$ K) and 304~{\AA} (chromosphere and transition region, $T=50\,000$~K). These channels were selected for their characteristic temperature to complement the ALMA data. HMI data provided photospheric line-of-sight magnetic field measurements. We used a cadence of 1 minute for AIA images and 45 seconds for HMI data. Both AIA and HMI data were preprocessed with routines from the Python \texttt{aiapy} software package, which is analogous to the usual IDL \texttt{aia\_prep} procedure.

For time profiles, we selected a circular region of a size large enough to cover the entire flaring region and at least the ALMA beam size ($\sim$60{\arcsec} in band 3 and $\sim$28{\arcsec} in band 6). The radius of 100{\arcsec} was used for the first four flares and 50{\arcsec} for the SOL2018-12-15 flare. Both the average intensity over this flaring region and the intensity of the brightest pixel in the region were measured in all ALMA and SDO/AIA images at all times. The position of peak intensity changes over time and in different wavelengths. While the average intensity is useful for the overall comparison of the ALMA and SDO/AIA intensity profiles and estimation of the total energy released, the peak intensity refers to spatially separated individual brightenings and better assesses the maximum increase of the brightness temperature.

The soft X-ray fluxes for all events were observed by the Geostationary Operational Environmental Satellites (GOES), specifically by the GOES-15 X-Ray Sensor (XRS), and were obtained from the National Oceanic and Atmospheric Administration (NOAA).

GONG H$\alpha$ images are from the Cerro Tololo Interamerican Observatory in Chile taken at the core of the H$\alpha$ line at 6562.8~{\AA} with a spatial resolution of 1{\arcsec} per pixel and 1 minute cadence.

We searched for bursts related to the analyzed flares in dynamic spectra recorded by the solar radio spectrometers e-CALLISTO in the 20 - 300 MHz range and Radio Solar Telescope Network (RSTN) GHz data, but found no events.

In the following intensity profile figures, gray shaded areas indicate the duration of each ALMA scan of the Sun, thus representing time uncertainty of the ALMA measurements. Average intensity of the circular region containing the flare is represented by a black curve while maximum intensity is depicted as a red curve. The mean ALMA intensity error was obtained from the variation of an intensity profile of a quiet Sun region.

The ALMA images relevant to the flare were further analyzed at the times indicated by vertical blue lines in the intensity profiles. To better isolate the flaring region, a difference image was made by subtracting a reference image. The reference image was usually selected as the ALMA image obtained just before the start of the flare. The effect of solar differential rotation was taken into account. Then the contours from the ALMA difference image were overlaid on the HMI magnetogram, the H$\alpha$ image from the Global Oscillations Network Group (GONG), and the AIA 304, 171, and 94~{\AA} filtergrams. This was done in order to find counterparts of the ALMA flare emission.

\section{Results}
\subsection{SOL2017-04-23 flare}

The SOL2017-04-23 flare is the only ALMA band 6 event in the analyzed set. The advantage of band 6 over band 3 is in twice the spatial resolution of the images, but at the expense of less temporal resolution due to the longer time required to scan the entire Sun. The beam size is 28.3{\arcsec}, and the pixel size is 3{\arcsec}. ALMA images were produced from PM01 antenna data at 230~GHz (1.3~mm).

The observed ALMA and SDO/AIA intensity profiles are shown in Fig. \ref{fig-lc-20170423} and compared with GOES XRS flux. The time of maximum millimeter brightness temperature coincides with the maxima in the other wavelengths. The average ALMA profile and the peak ALMA profile are similar except for the last measurement, where the average rises but the peak values stay constant. The reason for this is an increase in total radiation distributed over a larger area (compare the contours at 16:14 UT and 16:34 UT in Fig. \ref{fig20170423}). The average EUV profiles are similar to each other and to the GOES X-ray curve as well. 

The peak intensity profiles of the various wavelengths differ more from each other, especially in 171~{\AA} channel, where many multiple post-maximum peaks can be seen. They indicate strong brightenings originating from individual small areas (high loops).

 ALMA images of the flaring region are shown in Fig. \ref{fig20170423} and compared with HMI LoS magnetograms, GONG H$\alpha$ images, and AIA filtergrams with overlaid ALMA 100 - 500~K contours. H$\alpha$ and 304~{\AA} images suggest a two-ribbon flare scenario. The first image shown, (Fig. \ref{fig20170423}, 15:57), corresponds to the peak of the flare. The ALMA contours encompass the extent of the flare in H$\alpha$, 304 and 94~{\AA} images, and coincide with the polarity inversion line in the HMI magnetogram. The peak ALMA brightness in the flaring region, marked with a white "x", is located near the southern footpoint of a bright 94~{\AA} loop. The ALMA peak is also located near the top of a small loop visible in H$\alpha$, 304, and 171~{\AA}. 

A second ALMA component, peaking to the north, coincides with a small filament visible in H$\alpha$ absorption. The footpoint of a thin loop is visible at the same location in 304 and 171~{\AA}. A third ALMA source with significant brightness is located northwest (up right) of the main peak. The movies in H$\alpha$ and 304~{\AA} suggest that the third source is at the location of plasma downflow originating from a region near the second peak.

   \begin{figure}[h!]
   \centering
   \includegraphics[width=9.2cm]{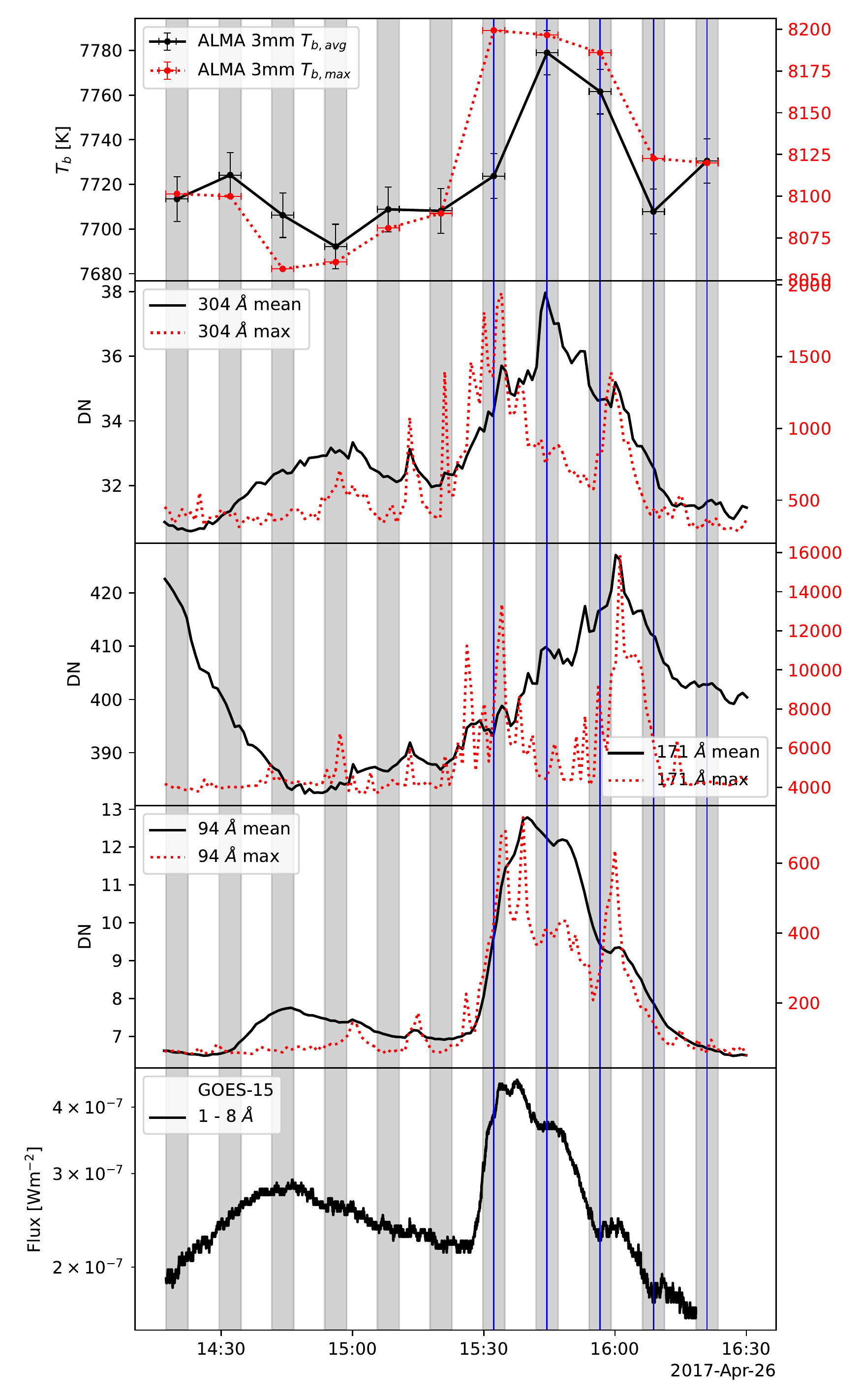}
   \caption{Intensity profiles for the SOL2017-04-26 flare. The
   colors and markings are the same as in Fig.
   \ref{fig-lc-20170423}.}
              \label{fig-lc-20170426}%
    \end{figure}

The next time shown, (Fig. \ref{fig20170423}, 16:14), is around the secondary flare peak, visible in 304 and 171~{\AA} average intensity profiles (Fig. \ref{fig-lc-20170423}). The ALMA peak brightness is shifted to the north, next to a small H$\alpha$ filament that became more pronounced in H$\alpha$ absorption. The ALMA peak location is consistent with the top of the H$\alpha$ filament but not with any feature at any other wavelength. However, an intensifying plasma flow can be observed in 304 and 171~{\AA} movies from the main ALMA location, with some of the material falling back to the surface near the location of the second ALMA bright region (on the right).

Finally, in the last ALMA image (Fig. \ref{fig20170423}, 16:34), the H$\alpha$ filament has intensified (darkened), and the ALMA contours now correspond well to its shape. The maximum ALMA brightness is situated at the southern footpoint of the filament. The area at the location of plasma inflow (to the right of the filament) has slightly brightened. The original ALMA peak located above the flare ribbons has also brightened and matches the shape of the hot flare loop visible in 94~{\AA}.

   \begin{figure*}[h!]
   \centering
   \includegraphics[width=18.5cm]{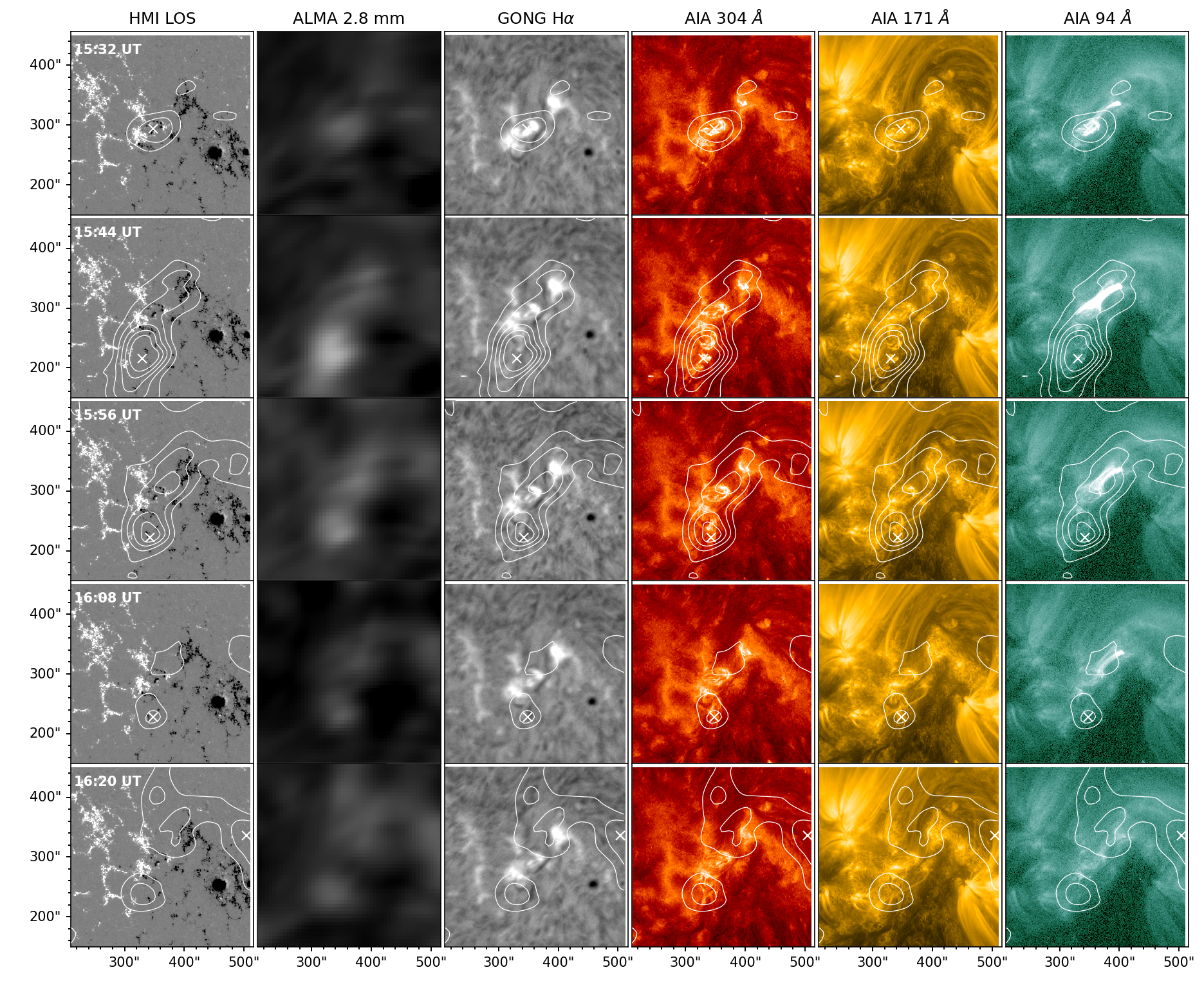}
   \caption{SOL2017-04-26 flare. ALMA contours outline five levels equidistantly in the 50 - 200~K range.}
              \label{fig20170426}%
    \end{figure*}

We note in conclusion that the millimeter emission of SOL2017-04-23 is associated with a bewildering number of features: 

- footpoint of hot (94~{\AA}) loop (main ALMA peak)

- top of cool loops in H$\alpha$, 304, and 171~{\AA}

- downflow region in H$\alpha$, 304, and 171~{\AA}

- top and footpoint of dark H$\alpha$ filament.

\subsection{SOL2017-04-26 flare}

The flare that occurred on 2017 Apr 26 was recorded by ALMA in band 3 in the spectral window centered at 107~GHz (2.8~mm). The beam size is 60{\arcsec} and pixel size is 6{\arcsec}. Artifacts from the antenna scanning pattern can be seen in ALMA images, especially in the difference images, somewhat affecting the observed shape of the flaring region.

The region-averaged ALMA time profile looks similar to 304~{\AA}, 94~{\AA}, and GOES, but the 171~{\AA} emission peaks  significantly later (Fig. \ref{fig-lc-20170426}). The pre-flare 171~{\AA} average intensity level was even higher than during the flare. This brightness originates from activity within large loops arching over the region. The ALMA peak intensity behaves similarly to the average ALMA intensity. The only difference is that peak curve maximum occurs before the average maximum, in line with the 304~{\AA} channel behavior.

 The ALMA peak in the first image at the time of the flare maximum coincides with the bright spot in 94~{\AA} (Fig. \ref{fig20170426}, 15:32). The region is also bright in H$\alpha$ and 304~{\AA} images, but not in the 171~{\AA} image. Most 171~{\AA} emission originates from high loops having many small brightenings that do not show in the ALMA image. These brightenings are visible in H$\alpha$ and 304~{\AA} images and contribute to the high temporal variability at these wavelengths. 
 Whereas the ALMA maximum brightness falls exactly on the 94~{\AA} peak in the flare rise phase, the ALMA peak shifts south along the neutral line in the following time steps. The ALMA contours outline the whole flaring region in the next time step (Fig. \ref{fig20170426}, 15:44). The peak brightness is at the location of the peak in 304~{\AA}. A new thin loop can be seen nearby in 171~{\AA}. The ALMA peak remains in the same place in the next two frames (Figs. \ref{fig20170426}, 15:56 - 16:08) where small filamentary structures are forming in 304~{\AA}, while nothing can be seen in 94~{\AA}. However, a secondary ALMA peak at 16:20 UT (Fig. \ref{fig-lc-20170426}) is located on two decaying 94~{\AA} flare loops in the northwest.
 
 Millimeter emission of the SOL2017-04-26 flare is associated with:
 
 - footpoint of a hot flare loop (94~{\AA})

 - small 171~{\AA} brightenings

 - decaying hot loops in 94~{\AA}.

\subsection{SOL2018-04-03 flare}

The SOL2018-04-03 flare was a short-lived two-ribbon event. The flare lasted less than 20 minutes, but evolution and post-flare aftermath were well recorded in 23 ALMA images at 95~GHz (3.2~mm). The beam size is 66{\arcsec} and pixel size is 6{\arcsec}.

   \begin{figure}[h!]
   \centering
   \includegraphics[width=9.2cm]{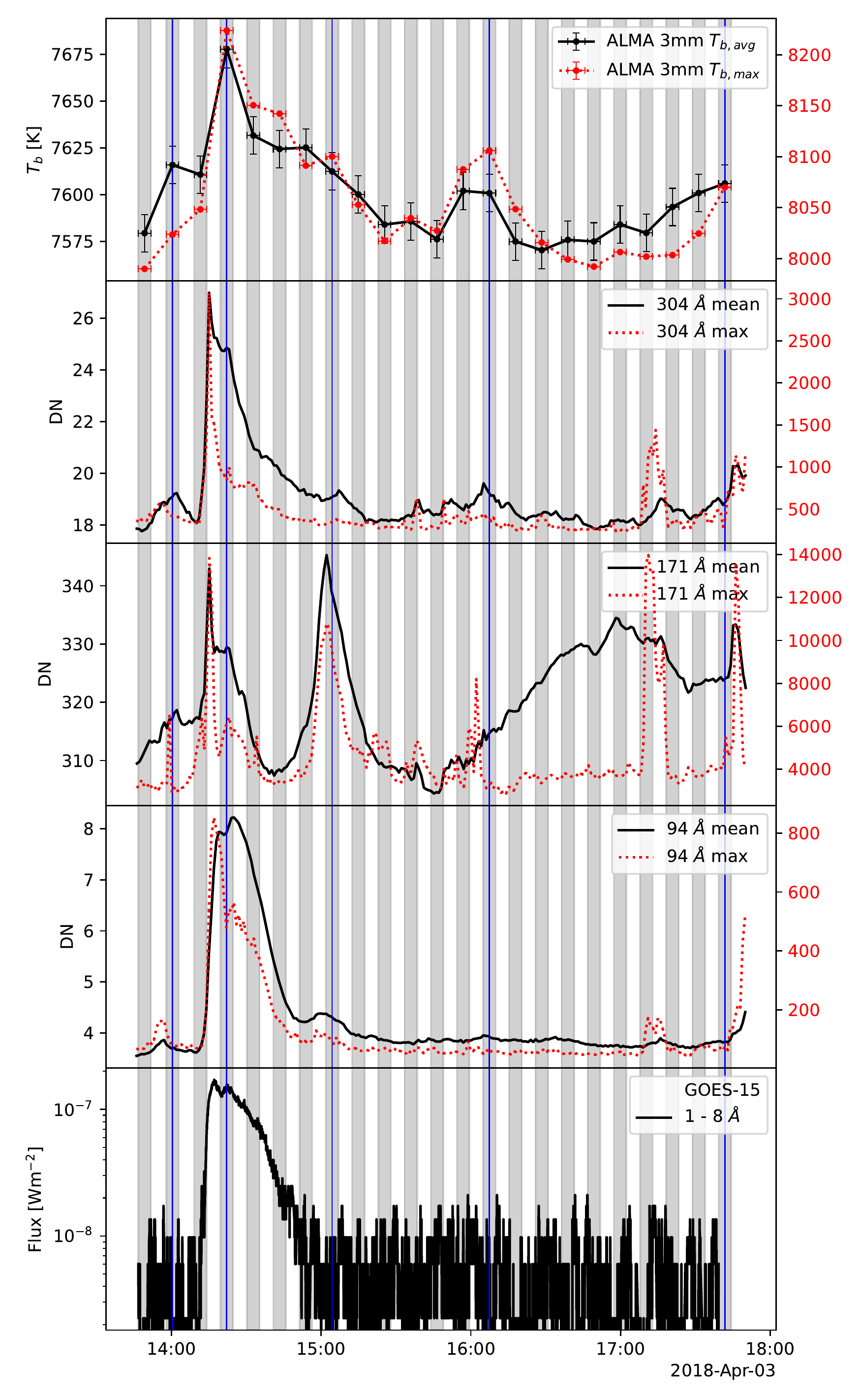}
   \caption{Intensity profiles for the SOL2018-04-03 flare. The
   colors and markings are the same as in Fig.
   \ref{fig-lc-20170423}}
              \label{fig-lc-20180403}%
    \end{figure}

The main ALMA peak seems to be delayed in time compared to all other channels. However, this may be the result of undersampling of the ALMA time profile, missing the peak in millimeter waves.  

\begin{figure*}[h!]
   \centering
   \includegraphics[width=18.5cm]{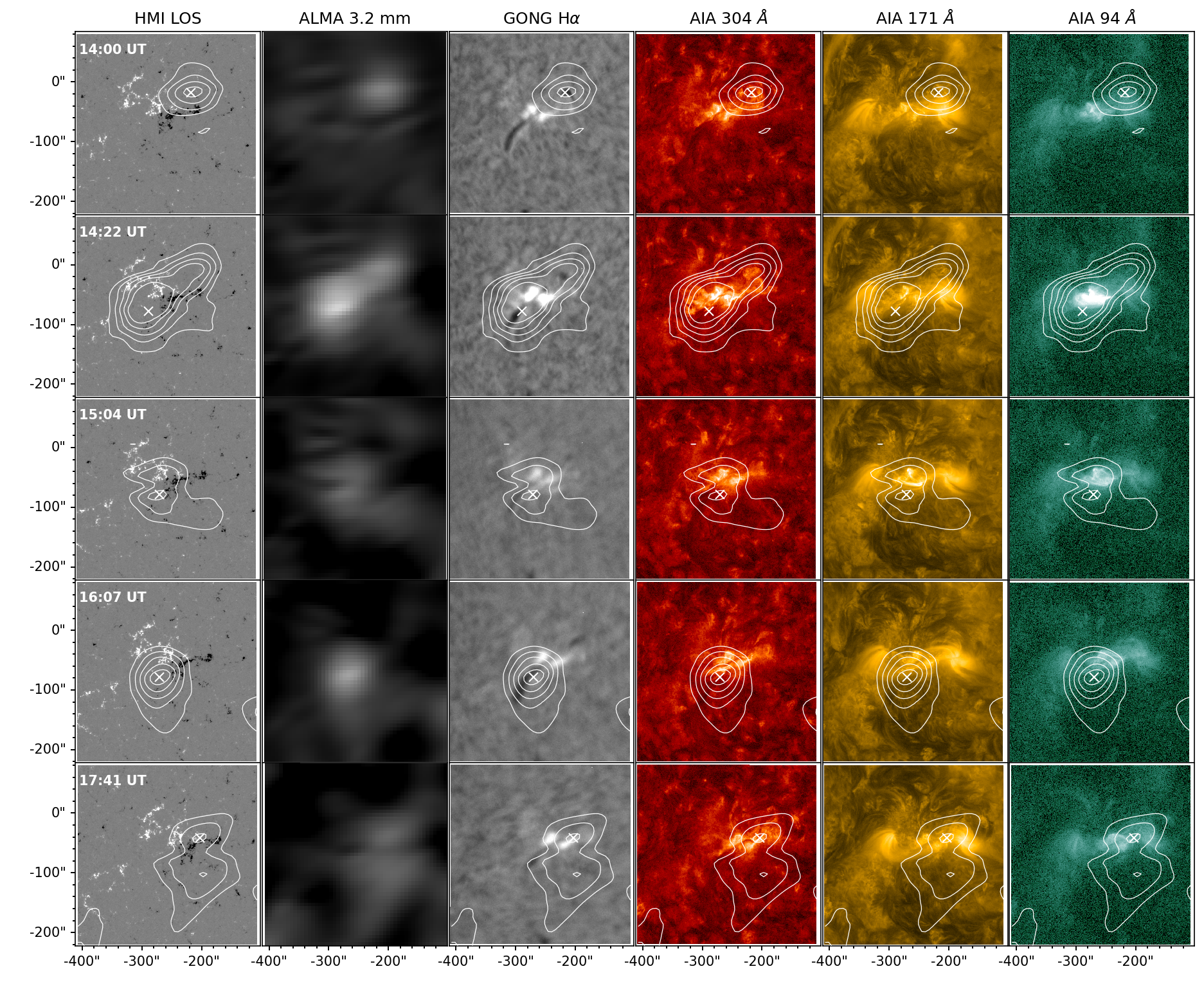}
   \caption{SOL2018-04-03 flare. ALMA contours outline five levels equidistantly in the 50 - 200~K range.}
              \label{fig20180403}%
    \end{figure*}

The preflare seen in the ALMA time profile (Fig. \ref{fig-lc-20180403}) coincides in space with a dark H$\alpha$ filament (Fig. \ref{fig20180403}, 14:00). Yet, the H$\alpha$ preflare brightening is related to emissions from the two ribbons located some 28{\arcsec} to the southeast. Interestingly, there is initially no visible ALMA enhancement around the prominent filament located southeast of the small one.  

At the peak of the ALMA flare (Fig. \ref{fig20180403}, 14:22), however, the ALMA maximum moves to the northern footpoint of the larger filament, and now also encompasses the flare ribbons and loops visible in 304 and 94~{\AA}. The smaller filament remains enhanced in the ALMA image. The ALMA emission follows in general the neutral line (Fig. \ref{fig20180403}, 14:22). At the same time, the H$\alpha$ filament shrinked in diameter and length.

Further in time, the ALMA peak remains on the larger filament which is now more pronounced. The prominent peaks in 171~{\AA} at 15:04 and 17:10 UT are caused by large loops leaving no traces in millimeter waves. The ALMA brightness slightly increases at 16:07, when plasma erupts from the active region and moves south along the major filament, as can be seen in 304~{\AA} movies. Finally, at 17:41 UT, the ALMA intensity peaks near the southern H$\alpha$ ribbon at a footpoint of a hot loop visible in 94~{\AA} emission.

The SOL2018-04-03 flare is an example where the ALMA peak emission is best associated with H$\alpha$ filaments. They are not steadily bright in millimeter emission, but become visible due to some activity related to the flare. Millimeter and 94~{\AA} emission are spatially not related except for a microflare in the late post-flare phase.

\subsection{SOL2018-04-19 flare}

\begin{figure}[h!]
   \centering
   \includegraphics[width=9.2cm]{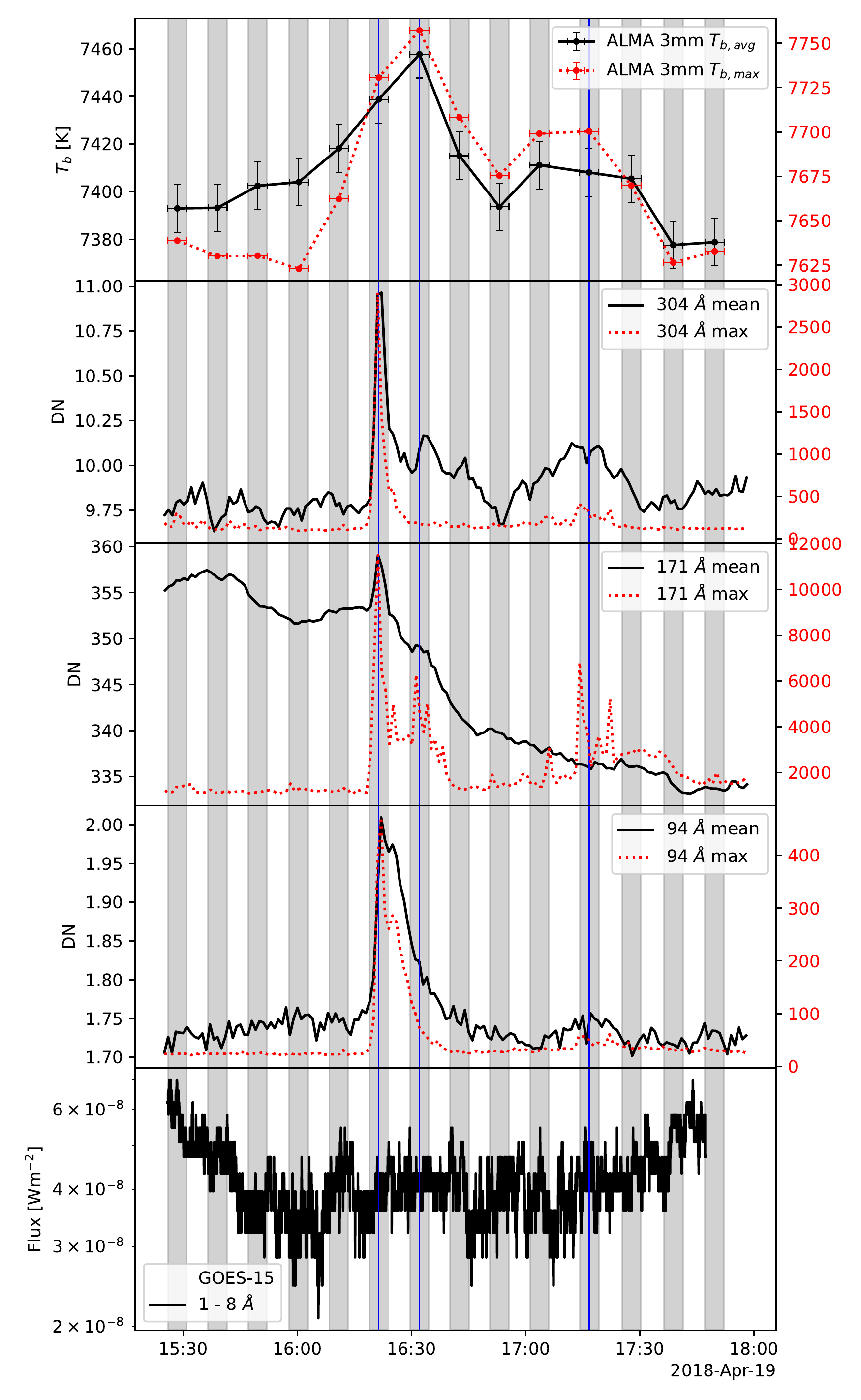}
   \caption{Intensity profiles for the SOL2018-04-19 flare. The
   colors and markings are the same as in Fig.
   \ref{fig-lc-20170423}}
              \label{fig-lc-20180419}%
    \end{figure}

\begin{figure*}[h!]
   \centering
   \includegraphics[width=18.5cm]{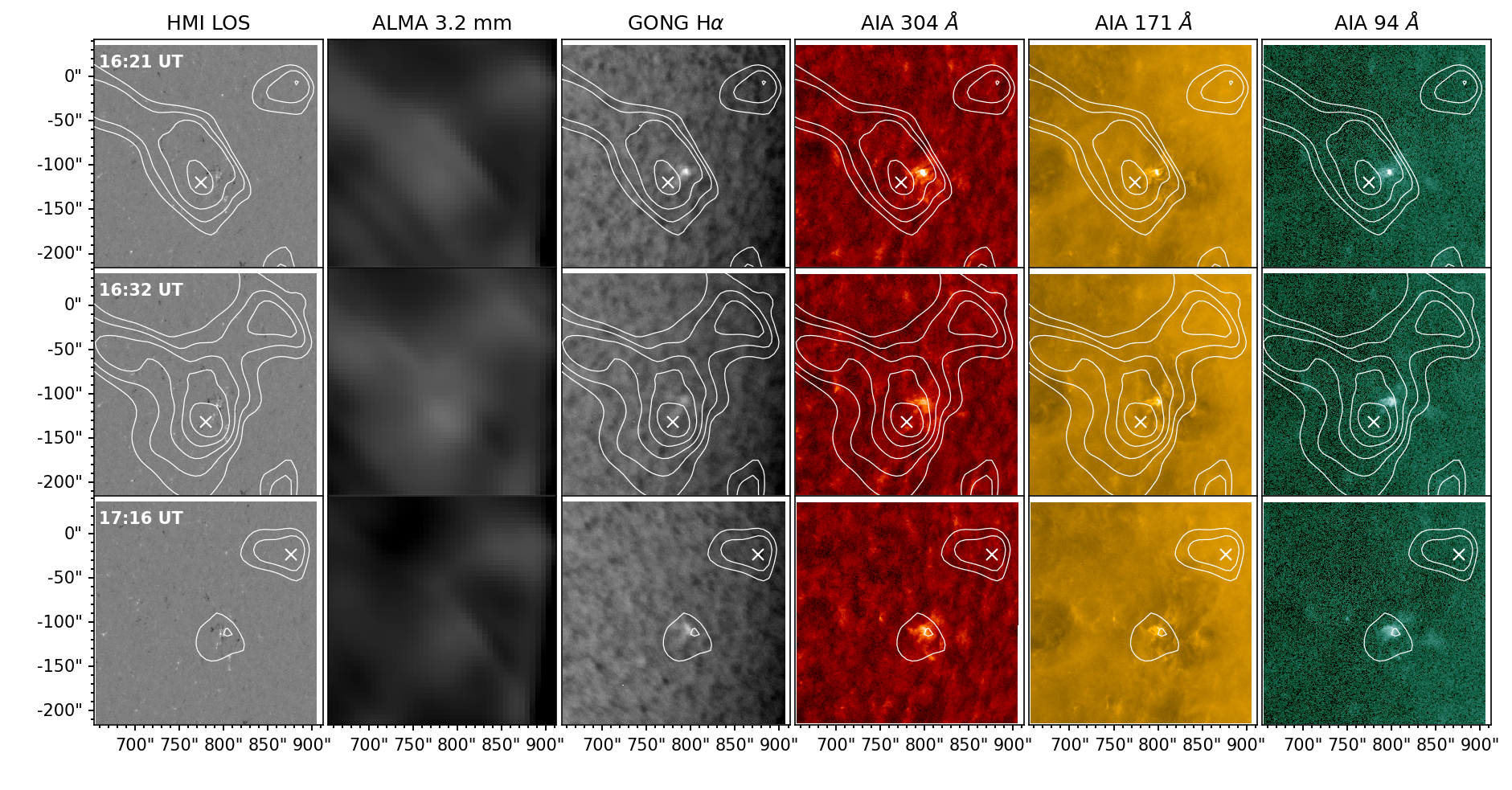}
   \caption{SOL2018-04-19 flare. ALMA contours outline five levels equidistantly in the 50 - 120~K range.}
              \label{fig20180419}%
    \end{figure*}

The SOL2018-04-19 flare was spotted while manually browsing ALMA images. GOES-15 did not detect it in soft X-rays (Fig. \ref{fig-lc-20180419}). Thus there is no entry in the HEK flare catalog. The flare is short and compact, and located near the western solar limb where the ALMA antenna scanning pattern  smears out sources diagonally. This is well noticeable in the background of Fig. \ref{fig20180419}.

The main peak in the EUV lines is sharp and precedes the ALMA peak by around 11 minutes, although the rise of ALMA intensity is visible even before. The ALMA peak coincides in time with a small peak in 304~{\AA} and 171~{\AA} about 11 minutes after their main peak. The subpeak is not visible in 94~{\AA}, indicating the absence of hot plasma. 

The main ALMA peak is located 25{\arcsec} southeast of the maxima in the line emissions (Fig. \ref{fig20180419}, 16:32). The ALMA images show a secondary peak northwest of the main peak. There is no corresponding brightening in the line emissions. This is the case even in the last image, when the secondary ALMA peak exceeded the primary peak (Fig. \ref{fig20180419}, 17:16).

The SOL2018-04-19 flare is a case where ALMA millimeter waves correlate with cool (H$\alpha$ and 304~{\AA}) lines, with the intermediate 171~{\AA} line, the hot 94~{\AA}, and the extremely hot bremsstrahlung emission (GOES) in time, but are displaced in space.

\subsection{SOL2018-12-15 flare}

   \begin{figure}[h!]
   \centering
   \includegraphics[width=9.2cm]{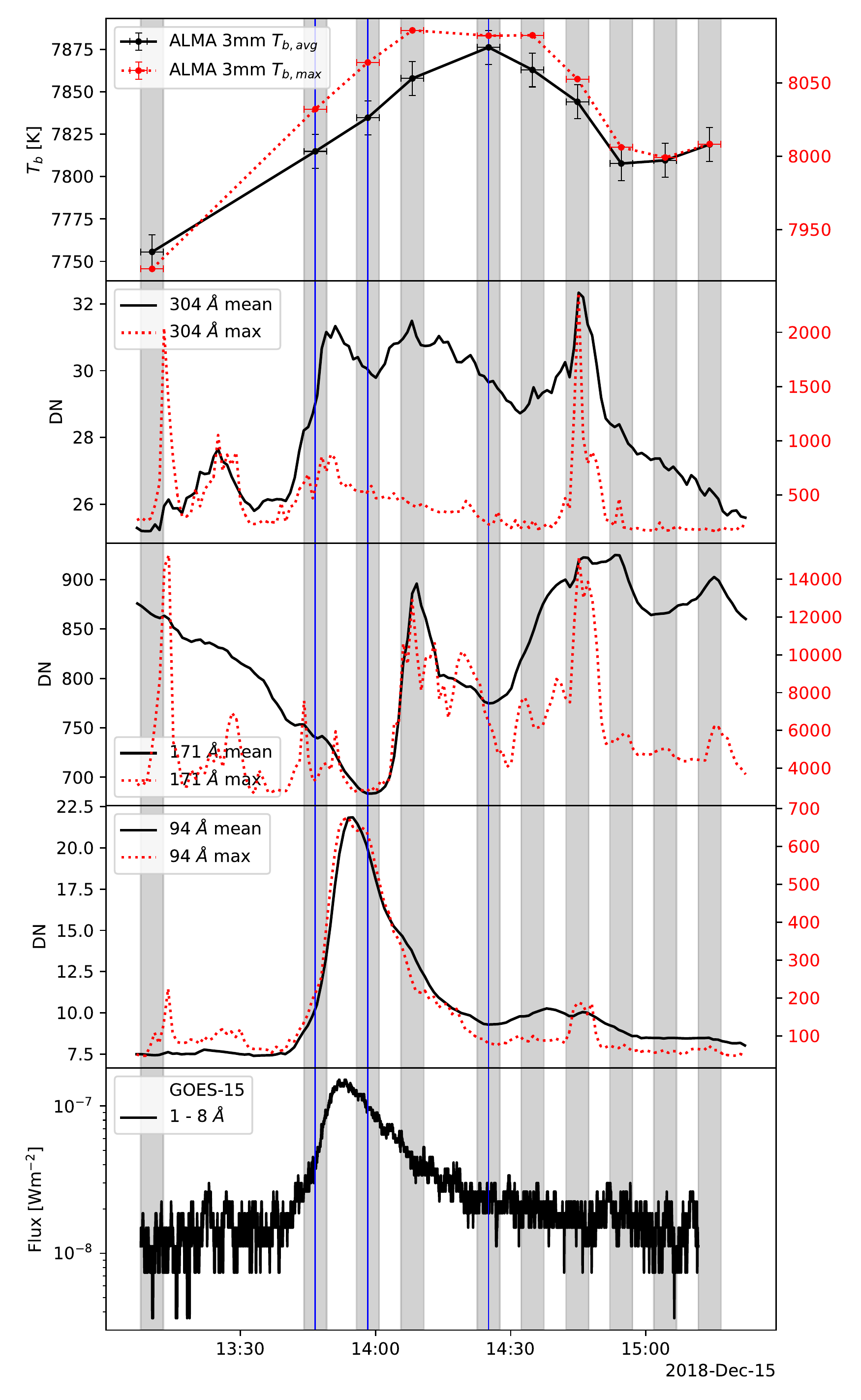}
   \caption{Intensity profiles for the SOL2018-12-15 flare. The
   colors and markings are the same as in Fig.
   \ref{fig-lc-20170423}}
              \label{fig-lc-20181215}%
    \end{figure}

\begin{figure*}[h!]
   \centering
   \includegraphics[width=18.5cm]{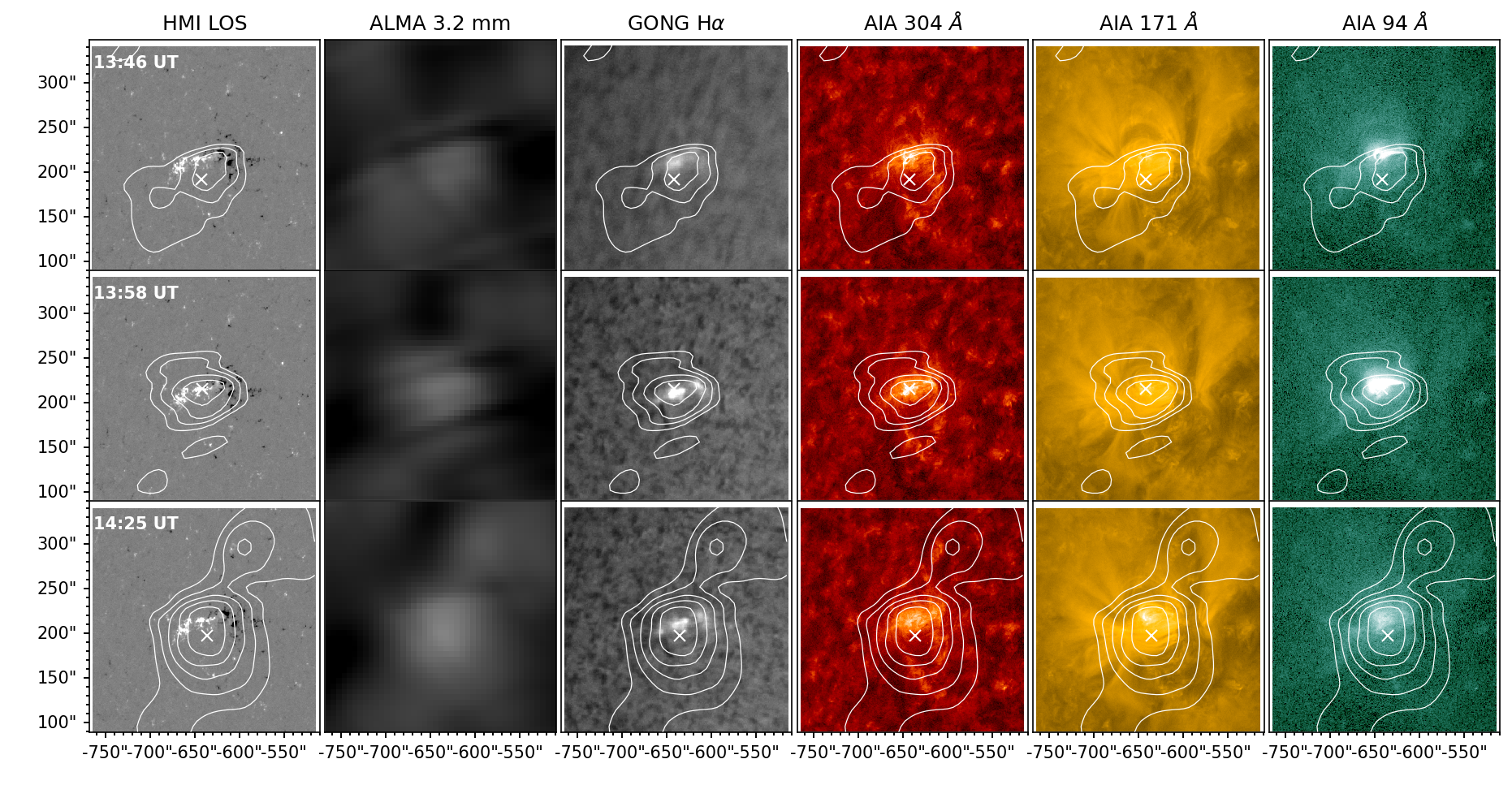}
   \caption{SOL2018-12-15 flare. ALMA contours outline five levels equidistantly in the 50 - 150~K range.}
              \label{fig20181215}%
    \end{figure*}

The weak SOL2018-12-15 flare occurred close to the limb. ALMA was observing at 95~GHz (3.2~mm). The observed intensity profiles are shown in Fig. \ref{fig-lc-20181215}. The ALMA intensity profile correlates best with 304~{\AA} average intensities, but not with the GOES X-ray and the 94~{\AA} profile, which are similar to each other. The flare reached its peak somewhat earlier in 304~{\AA} relative to the millimeter waves, but much later in 171~{\AA}. The 171~{\AA} emission is dominated by large loops arching over the active region (Fig.\ref{fig20181215}, 13:58). The averaged profile has a minimum intensity at the time of the ALMA flare peak (13:58 UT) due to dimming of some of the loops.

The brightening in millimeter waves encompasses most of the active region as seen in the magnetogram, but peaks 25{\arcsec} farther south where little emission in H$\alpha$, 304~{\AA}, 171~{\AA}, and 94~{\AA} originates (Fig.\ref{fig20181215}, 13:46 and 14:25). Only at the peak time (13:58 UT) of the X-ray and 94~{\AA} intensities, the peak position of all line emissions coincide with the peak in millimeter waves. 

The SOL2018-12-15 flare is another example of weak millimeter emission with little relation to the emissions in H$\alpha$, 304~{\AA}, 171~{\AA}, 94~{\AA}, and X-rays.

\section{Discussion}

\begin{table*}

\caption{List of features in other wavelengths coincident with emission in millimeter ALMA images for each flare.}            
\label{tabSummary}      
\centering                          
\begin{tabular}{l l l p{10cm}}        
\hline\hline                 
 Date & Time UT  &  Peak source  &  Coincides with \\
 &   &  in mm image &  \\    
\hline                        
2017-04-23 & 15:57 &  primary  &  along neutral line in HMI, near the top of a small loop in H$\alpha$, 304 and 171~{\AA}, footpoint of a 94~{\AA} loop \\
&  &  secondary  &  top of H$\alpha$ filament, thin loop footpoint at 304 and 171~{\AA} \\
&  &  tertiary  &  plasma downflow in H$\alpha$, 304~{\AA} \\
& 16:14 &  primary  &  near the top of H$\alpha$ filament, plasma outflow in 304 and 171~{\AA} \\
& 16:34 &  primary  &  footpoint of a dark H$\alpha$ filament \\
\hline   
2017-04-26 & 15:32 & primary & footpoint of a flare loop in 94~{\AA}, 
bright in H$\alpha$ and 304~{\AA}, 
nothing in 171~{\AA}\\
& 15:44-15:56 & primary & along neutral line in HMI, newly forming filament in H$\alpha$, peak in 304~{\AA}, top of loop in 171~{\AA}, nothing in 94~{\AA}\\
& 16:20 & secondary & decaying flare loops in 94~{\AA}\\
\hline
2018-04-03 & 14:00 & primary & H$\alpha$ filament\\
 & 14:22 & primary & along neutral line in HMI, H$\alpha$ filament footpoint, thin loops in 304 and 171~{\AA}, near the hot loop top in 94~{\AA} \\
 & 14:22-16:07 & primary & H$\alpha$ filament footpoint\\
 & 17:41 & primary & H$\alpha$ ribbon, footpoint of a hot loop in 94~{\AA}\\
\hline
2018-04-19 & 16:21-16:32 & primary & nothing (displaced from the flaring region) \\
\hline
2018-12-15 & 13:46 & primary & loops in 171~{\AA}, nothing elsewhere\\
 & 13:58 & primary & top of the loop in H$\alpha$ and 304~{\AA}, footpoint of the flaring loop in 94~{\AA}\\
\hline                                   

\end{tabular}
\end{table*}

\begin{table}
\caption{NOAA number of the flare active region, peak pixel brightness temperature $T_b$ at wavelength $\lambda$ and peak pixel brightness temperature difference $\Delta T_b$ from the background level.}             
\label{tabFlareMax}      
\centering                          
\begin{tabular}{c c c c c}        
\hline\hline                 
 Flare & NOAA & $\lambda$ & Peak $T_b$ &  Peak $\Delta T_b$\\
 & region & [mm] & [K]   &  [K]  \\    
\hline                        
SOL2017-04-23 &   N/A      & 1.3 & 7163  &  563  \\
SOL2017-04-26 &  12652  & 2.8 & 8199  &  263  \\
SOL2018-04-03 &  12703  & 3.2 & 8224  &  321  \\
SOL2018-04-19 &   N/A     & 3.2 & 7757  &  142  \\
SOL2018-12-15 &  12731  & 3.2 & 8086  &  194  \\
\hline                                   

\end{tabular}
\end{table}

Significant millimeter radiation was found in all 4 flares selected from the HEK catalogue. In the event selected from ALMA data, brightenings in line emission in H$\alpha$, 304~{\AA}, 171~{\AA}, and 94~{\AA} are associated. This strongly confirms a general correlation in time between the different radiations originating from very hot, hot, intermediate and cool plasma. The comparison of the images, on the other hand, shows a surprising variety of associations with spatial features. A summary of features cospatial with the millimeter emission for each flare is given in Table~\ref{tabSummary}.

In both ALMA bands 3 and 6, the ALMA flare emission occurs mainly above the neutral line observed in HMI magnetograms. Good examples are Fig. \ref{fig20170423} at 15:57, Fig. \ref{fig20170426} at 15:44 and Fig. \ref{fig20180403} at 14:22. However, this is not always the case and millimeter emission may correspond to different features such as 

- H$\alpha$ active filaments (Fig. \ref{fig20170423} at 16:14 and 16:34, Fig. \ref{fig20180403} at 14:00, 14:22, and 16:07) 

- 94 \AA~ (hot) loop tops (Fig. \ref{fig20170426} at 15:56) or footpoints (Fig. \ref{fig20170423} at 15:57), 

- post-flare loops (Fig. \ref{fig20170426} at 16:08)  

- occasionally to no features in other wavelengths at all (Fig. \ref{fig20180419} at 17:16, secondary peak in Fig. \ref{fig20180403} at 17:41)

Millimeter emission originating from flaring loop tops and footpoints was found previously at 86 GHz with the BIMA interferometer \citep{Silva1996,Silva1998}. They estimated that the major part (around 80\%) of the emission in the flare gradual phase came from the top of a loop as free-free radiation emitted by hot plasma. \citet{Kundu2000} reported a clear detection of both footpoints of a loop at $\lambda$=3 mm.

It is interesting to note that millimeter flare emission sometimes corresponds very well with filaments, particularly to the activity within them (e.g. SOL2018-04-03). \citet{Rutten2017} predicted that the appearance of the Sun at ALMA wavelengths will be similar to the one in H$\alpha$ with good dark-dark correspondence. \citet{Brajsa2021} did find dark filaments at 3 millimeters. On the other hand, \citet{daSilvaSantos2022} found some of the dark threads visible in the AIA 304~{\AA} and Mg II resonance lines to have dark counterparts at 1.25~mm, but their visibility varied significantly across the filament both spatially and temporally. Here we report occasional dark-bright correspondence.

Two kinds of time profiles were extracted from the images. The average over the active region, shown with a black solid curve, corresponds to what a low resolution telescope would measure. In a second time profile, represented by a red dashed curve, the peak of the brightest pixel in the area is shown. The second curve indicates small, individual peaks of emission. 

In all time profiles, the peak pixel curves have more maxima and shorter maxima than the spatially averaged curves. It indicates that the millimeter flare emission originates from different peaks within the active region. The effect of a multitude of individual peaks is more pronounced in the SDO/AIA lines, especially at 171~{\AA}.

The peak pixel intensity measured from the intensity profiles and from the difference ALMA images is given in Table~\ref{tabFlareMax}. The largest increase is found in the band 6 flare with a value of $\Delta T_b = 563$~K above the pre-flare level, while in band 3 the values vary between 140 and 320~K. This corresponds to 500 -- 1200~K above the quiet Sun level. These values are lower limits since the maximum intensity is spread by the antenna beam for compact, unresolved sources.

To investigate statistically the observed similarity of the ALMA and SDO/AIA EUV intensity curves, we calculated the Pearson correlation coefficient between ALMA and EUV, average and maximum, time profiles. The results are listed in Table \ref{tabCorr}. 

The correlation between average ALMA profile and AIA 94 and 304~{\AA} channels is quite good. In three cases, the 304~{\AA} channel, indicating cool plasma, has the highest correlation coefficient. This corresponds to earlier reports about spatial best correlation with the 304~{\AA} line among all EUV AIA lines (\citealt{White2017, Brajsa2018AA}).

The 94~{\AA} line, originating from hot plasma, has a slightly higher correlation of the remaining two cases (Table \ref{tabCorr}). However, the 171~{\AA} line, emitted by plasma of intermittent temperature in coronal loops, has a much lower correlation coefficient, or even an anti-correlation. A similar trend with low 171~{\AA}, and high 94 and 304~{\AA} correlations is present in peak intensity profile. However, there are two exceptions where 171~{\AA} peak emission has a similarly high correlation as 94 and 304~{\AA} channels. The first one is SOL2017-04-23 flare where peak pixel 171~{\AA} curve follows the shape of the ALMA peak and average profiles. The second one is SOL2018-04-19 flare where the high 171~{\AA} to ALMA correlation is due to the secondary peak(s).

The observed similarity between 304~{\AA}, 94~{\AA} and ALMA time profiles suggests that the millimeter brightness in flares originates mainly from chromospheric phenomena, but occasionally also from hot and dense flare components. Geometry could have a role in the observed correlations as well since the last two flares from Table~\ref{tabCorr}, with generally the lowest correlation from the set, are also closest to the solar limb.

The accuracy of the contours and the location of maximum intensity is better than the ALMA beam width. The confidence interval depends on statistical significance of the peak above background and the number of samplings of the single-dish beam by the double-circle
scanning pattern. This pattern consists of minor circles with a radius of 600{\arcsec}, whose centers move steadily in a major circle around the center of the Sun at a distance of 600{\arcsec}. Each new minor circle is shifted along the major circle by a defined value called the sampling length \citep{White2017}. In the worst case scenario when minor circles fall on top of each other, the largest distance of the sampled points is less than or equal to the sampling length. So, the beam is at least sampled at 3 times the beam resolution, and in many places like near the disk center even more, giving a minimum resolution of 10{\arcsec} in band 6 and 20{\arcsec} in band 3. The location of the peak can be determined even more precisely by centroiding.

The flares of this work were not detected in hard X-rays
 and microwaves. Using the time derivative as a proxy for the hard
 X-ray flux, we estimate the impulsive phase of the flare by the
 rise phase of the soft X-rays from beginning to peak. In the four 
 flares with significant soft X-rays, the ALMA millimeter
 intensity is consistent with peaking after the impulsive phase.
 Flare SOL2017-04-23 (Fig.~\ref{fig-lc-20170423}), however, is 
 also consistent with a  maximum during the impulsive phase, and 
 the peak intensity curve (red) of flare 
 SOL2017-04-26 (Fig.~\ref{fig-lc-20170426}) has its maximum 
 clearly during the impulsive phase. Thus, >100 GHz flares 
 appear to be  mostly gradual phase phenomena with occasional 
 impulsive phase emission.

\begin{table}
\caption{Pearson correlation coefficients between intensity measurements from ALMA and different AIA channels. {\it Top:} Flare region averaged, {\it bottom:} peak pixel of flare region.}             
\label{tabCorr}      
\centering                          
\begin{tabular}{l c c c c c}        
\hline\hline                 
 Intensity & Date  &  N  &  94~{\AA}  & 171~{\AA} &  304~{\AA} \\
 profile &   &   &   &  & \\    
\hline                        
\multirow{5}{0pt}{Average} & 2017-04-23 &  9  &  0.797  &  0.598  &  0.823 \\
& 2017-04-26 &  11  &  0.777  &  0.582  &  0.746 \\
& 2018-04-03 &  23  &  0.799  &  0.045  &  0.882 \\
& 2018-04-19 &  14  &  0.721  &  0.391  &  0.613 \\
& 2018-12-15 &  10  &  0.330  &  -0.175  &  0.757 \\
\hline                                   
\multirow{5}{0pt}{Peak pixel} & 2017-04-23 &  9  &  0.939 & 0.924 & 0.938 \\
& 2017-04-26 &  11  &  0.927 & 0.665 & 0.776\\
& 2018-04-03 &  23  &  0.771 & 0.064 & 0.713 \\
& 2018-04-19 &  14  &  0.495 & 0.599 & 0.421\\
& 2018-12-15 &  10  &  0.373 & 0.367 & 0.168\\
\hline                                   

\end{tabular}
\end{table}

\section{Conclusions}

Full-disk ALMA imaging has yielded the first spatially complete solar flare observations in millimeter wavelengths. The field of view includes the whole disk. We found a field of 250{\arcsec}$\times$250{\arcsec} necessary for a complete overview over the activity in millimeter waves taking place in an active region during a solar flare. It significantly exceeds the field of view of ALMA in interferometric mode. Single-dish observations thus complement interferometric observations in an important way. 

Most surprising is the fact that several phenomena in both hot and cold plasma lead to enhanced brightness in millimeter waves. Although there is no evidence contradicting the assumption that the emission process is thermal free-free emission, it seems that the previous scenario of a chromospheric hot spot heated by a precipitating electron beam is too simplistic. 

In addition to footpoints of hot loops, millimeter flare emission was found to be associated with activated H$\alpha$ filaments, impact points of plasma motions, post-flare loops, and hot loop tops. Surprisingly, we found cases where no feature in H$\alpha$, 304~{\AA}, 171~{\AA}, and 94~{\AA} was visible at the position of a millimeter wave emission peak. 

In this work, we focused on comparing flare intensity profiles and coincident spatial sources at different wavelengths. More information about the flares, such as temperature and density of the emitting plasma, could be inferred from comparison with simultaneous soft X-ray observations. Its analysis would exceed the scope of this paper and it is planned for future work. The detected flares were too small for significant hard X-ray and microwave emissions.

Higher temporal and spatial resolution are needed to gain a better insight into the properties of flares at millimeter wavelengths. Both problems can be solved by observing flares using ALMA interferometric mode, however, this is difficult due to the ALMA scheduling constraints unable to wait for a flare. The field of view in interferometric mode is small and a lot of luck is needed to point at the right location and at the right moment. A possible way to achieve better temporal resolution is to use the recently implemented total power regional mapping in single-dish mode. There is also a third possibility in the future to develop more advanced data analysis methods that could go beyond the ALMA full-disk single-dish images presented here. The same data may be much better exploited using specific characteristics and redundancies present in the data.

\begin{acknowledgements}
This work has been supported by the Croatian Science Foundation under the project 7549 "Millimeter and sub-millimeter observations of the solar chromosphere with ALMA". 

It has also received funding from the Horizon 2020 project SOLARNET (824135, 2019–2022).

This paper makes use of the following ALMA data: ADS/JAO.ALMA\#2016.1.01129.S, ADS/JAO.ALMA\#2016.1.00070.S, ADS/JAO.ALMA\#2017.1.00072.S, ADS/JAO.ALMA\#2017.1.01138.S, ADS/JAO.ALMA\#2018.1.01879.S.
 ALMA is a partnership of ESO (representing its member states), NSF (USA) and NINS (Japan), together with NRC (Canada), MOST and ASIAA (Taiwan), and KASI (Republic of Korea), in cooperation with the Republic of Chile. The Joint ALMA Observatory is operated by ESO, AUI/NRAO and NAOJ.

SDO/AIA data courtesy of NASA/SDO and the AIA, EVE, and HMI science teams. 

This work utilizes GONG data obtained by the NSO Integrated Synoptic Program, managed by the National Solar Observatory, which is operated by the Association of Universities for Research in Astronomy (AURA), Inc. under a cooperative agreement with the National Science Foundation and with contribution from the National Oceanic and Atmospheric Administration. The GONG network of instruments is hosted by the Big Bear Solar Observatory, High Altitude Observatory, Learmonth Solar Observatory, Udaipur Solar Observatory, Instituto de Astrofísica de Canarias, and Cerro Tololo Interamerican Observatory. 

We are grateful to the GOES team for making the data publicly available.

We acknowledge the use of the ALMA Solar Ephemeris Generator \citep{skokic2019seg}.

This research used version 3.1.4 \citep{sunpy_314} of the SunPy open source software package \citep{sunpy2020}. 
This research used version 0.6.4 of the aiapy open source software package \citep{Barnes2020}.
This research made use of Regions, an Astropy package for region
handling \citep{regions}.

\end{acknowledgements}

%
%

 \bibliographystyle{aa} 
 \bibliography{alma_sd_flares} 

\end{document}